\documentclass[aps,amsmath,superscriptaddress,nofootinbib]{revtex4-1}
\pdfoutput=1
\usepackage{bm}
\usepackage{epsfig}
\usepackage{amsfonts}
\usepackage{bbm}
\usepackage{color}

\def\bnslash{\bar n\!\!\!\slash}

\def\OMIT#1{}

\newcommand{\nn}{\nonumber}

\newcommand{\bea}{\begin{eqnarray}}
\newcommand{\eea}{\end{eqnarray}}

\newcommand{\beq}{\begin{equation}}
\newcommand{\eeq}{\end{equation}}

\begin{document}


\title{Probing Quarkonium Production Mechanisms with Jet Substructure} 

\author{Matthew Baumgart\footnote{Electronic address: baumgart@cmu.edu}}
\affiliation{Department of Physics, 
        Carnegie Mellon University,
	Pittsburgh, PA 15213
	\vspace{0.2cm}}

\author{Adam K. Leibovich\footnote{Electronic address: akl2@pitt.edu}}
\affiliation{Pittsburgh Particle Physics Astrophysics and Cosmology Center (PITT PACC)\\ Department of Physics and Astronomy, 
         University of Pittsburgh,
	Pittsburgh, PA 15260
	\vspace{0.2cm}}
	
\author{Thomas Mehen\footnote{Electronic address: mehen@phy.duke.edu}}
\affiliation{Department of Physics, 
	Duke University, 
	Durham,  NC 27708
	\vspace{0.2cm}}

\author{Ira Z. Rothstein\footnote{Electronic address: izr@andrew.cmu.edu}}
\affiliation{Department of Physics, 
        Carnegie Mellon University,
	Pittsburgh, PA 15213
	\vspace{0.2cm}}

\date{\today\\ \vspace{1cm} }


\begin{abstract}
We use fragmenting jet functions (FJFs) in the context of quarkonia to study the production channels predicted by NRQCD ($^3S_1^{(1)}, {}^3S_1^{(8)}, {}^1S_0^{(8)}, {}^3P_J^{(8)}$).  We choose a set of FJFs that give the probability to find a quarkonium with a given momentum fraction inside a cone-algorithm jet with fixed cone size and energy.  This observable gives several lever arms that allow one to distinguish different production channels. In particular, we show that at fixed momentum fraction the individual production mechanisms have distinct behaviors as a function of the the jet energy. As a consequence of this fact, we arrive at the robust prediction that if the depolarizing ${}^1S_0^{(8)}$ matrix element dominates, then the gluon FJF will diminish with increasing energy for fixed momentum fraction, $z$, and $z \, >$ 0.5.

\end{abstract}

\maketitle

\newpage


 \section{Introduction}
Nonrelativistic QCD (NRQCD) is an effective field theory  \cite{NRQCD} for quarkonium that reproduces full  QCD as an expansion in the relative velocity, $v$, of the heavy quark and antiquark. 
This theory  has been used to study both the decay and production of these bound states \cite{Bodwin:1994jh}. Its predictive power is predicated on our knowledge of a set of non-perturbative matrix elements that must be extracted
from the data. 
 In the case of $J/\psi$ or $\Upsilon$ production there are four such matrix elements that must be fit at leading order, and thus predictions
have mainly been limited to shapes of spectra. NRQCD predictions at NLO in the coupling  have been compared to the world data on $J/\psi$ production in Refs.~{\cite{Butenschoen:2011yh,Butenschoen:2012qr}. The $\chi^2$/d.o.f.~of 4.42\footnote{This $\chi^2$ is based on an analysis in which feed down from higher charmonia is ignored. Accounting for these contributions reduces the $\chi^2$ slightly to 3.74.} found in Ref.~\cite{Butenschoen:2011yh} is higher than one would hope for, but not unexpected given large theoretical uncertainties.

Thus, it is perhaps fair to say that we cannot yet claim that 
NRQCD is correctly describing quarkonium production with unqualified success. 
 In particular, one prediction  \cite{Cho:1994ih} of the theory
is that, at asymptotically large transverse momentum,  the $^3S_1$ state  ($J/\psi$ or $\Upsilon$) should be purely transverse at leading order. At present the data  in both the charm and bottom
sector do not see this trend \cite{Braaten-Russ} though the error bars are large, especially in the bottom system.  Furthermore the various experiments are not in agreement. 

It is important to appreciate that concluding that NRQCD is  ``wrong", in any sense, is equivalent to
saying that QCD does not properly describe these states. If NRQCD predictions for large $p_T$ production are not agreeing with the data, and we assume that the data is correct, then the only logical alternatives are:
(1) the velocity and/or $\alpha_s$ expansions are not converging,  (2) the fragmentation approximation, along with
its expansion in $m_Q/p_T$, is wrong, either due to the failure of factorization or the presence of anomalously large power corrections.  Let us consider each of these possibilities in turn. The perturbative corrections  in $\alpha_s(2m_Q)$ to the
fragmentation function were found to be small \cite{Beneke}. The possibility that the velocity power counting could
not apply to the charmed system \cite{Beneke2,FLR} is certainly a viable option, though the velocity expansion seems to work relatively
well for the decay processes \cite{Brambilla}. Moreover, one would expect for the bottom system that
the velocity expansion should converge nicely. It is possible that factorization is breaking down in the production
processes, as all such proofs, at least within the confines of SCET, are lacking a treatment of the factorization breaking
``Glauber mode". Nonetheless, given the success of semi-inclusive predictions in light hadronic systems, it would
be surprising to see a failure in the case of quarkonium.

A  more conservative guess would be that there is nothing wrong with the theory, but perhaps the values of extracted matrix elements are sufficiently inaccurate as to change the nature of the polarization prediction. For example, the magnetic spin flip operator could be anomalously large. In any case, to get a better handle on the situation we must improve our quantitative understanding
of the various production channels associated with the aforementioned matrix elements. The purpose of this paper is to introduce a new tool that will allow for a new extraction of these matrix elements by studying the characteristics of jets within which the quarkonium reside.

 \section{The Fragmenting Jet Function (FJF)} 
 
 Power counting dictates that at asymptotic values for  $p_\perp\gg m_Q$, quarkonia should be produced by single parton fragmentation.\footnote{In intermediate ranges of $p_\perp$ double-parton fragmentation 
 should dominate \cite{Sterman,FLMR1,FLMR2}. The phenomenology of double-parton fragmentation
 has yet to be performed.}  Since the parton initiating the fragmentation is a colored object, the quarkonium will be produced in association with light hadrons. In this paper we will consider  a $J/\psi$\footnote{The results will apply for the $\Upsilon$ as well. Of course the matrix elements will be different but most of the calculations
 in this paper are normalized such that the result is independent of the matrix element.
Thus, when we use the term $J/\psi$ we really mean the generic $^3S_1$ state.}
  produced  within a jet of energy $E$ and  cone size $R$, in which the $J/\psi$ carries a fraction of the jet energy, $z$. In this situation, a generic cross section is determined by the convolution of a hard and soft function 
(and possibly other jet functions, if there are other jets detected in the final state) multiplied by a quantity known as the fragmenting jet function (FJF),  first introduced in Ref.~\cite{Procura:2009vm} and further studied in Refs.~\cite{Liu:2010ng,Jain:2011xz,Jain:2011iu,Procura:2011aq,Jain:2012uq}. These papers focused on FJFs for light hadrons such as pions. FJFs for particles with a single heavy quark are studied in Ref.~\cite{Bauer:2013bza}. We show that the FJFs for gluon and charm quark jets containing a $J/\psi$ can be calculated in terms of a set of NRQCD long-distance matrix elements (LDME).  In our calculations the relevant LDMEs are: $\langle {\cal O}^{J/\psi}(^3S_1^{(1)})\rangle$, 
$\langle  {\cal O}^{J/\psi}(^1S_0^{(8)})\rangle$, $\langle  {\cal O}^{J/\psi}(^3S_1^{(8)})\rangle$,  and $\langle  {\cal O}^{J/\psi}(^3P_0^{(8)})\rangle$. 
The spectroscopic notation  indicates the quantum numbers of the heavy quarks prior to hadronization.
We show that the contribution
to the FJF from each of these mechanisms depends differently on
$z$ and $E$ and can thus be used to extract the LDME. Our results could easily be extended to jets  containing  other quarkonia states.

Since there are many observables associated with jets (angularities~\cite{Berger:2003iw}, broadening~\cite{Rakow:1981qn}, jet shape~\cite{Ellis:1992qq}, N-subjettiness~\cite{Thaler:2010tr}, etc.), one can generate a very large number of new tests of the NRQCD factorization formalism  by applying jet physics techniques to the study of quarkonia produced within jets. Furthermore, studying high $p_\perp$ quarkonia produced within jets avoids some of the potential theoretical pitfalls that could plague tests of the NRQCD factorization formalism at small $p_\perp$. At the highest $p_\perp$ available, we expect factorization to hold  up to corrections which scale as $m_Q/p_\perp$, and  furthermore the $\alpha_s$ expansion should be well behaved.  

\subsection{Operator Definitions}

We first briefly review the properties of the FJF~\cite{Procura:2009vm,Liu:2010ng,Jain:2011xz,Jain:2011iu,Procura:2011aq,Jain:2012uq}. 
We can consider many different production processes with a quarkonium inside a jet.  As an example, consider the two-jet cross section where one of the jets contains an identified $J/\psi$.  The factorization theorem \cite{Procura:2009vm} for the production cross section for a jet with energy $E$, cone size $R$, and a $J/\psi$ with energy fraction $z$ in a $pp$ collision is schematically of the form
\beq\label{fact}
\frac{d^2\sigma}{dE\, dz} = \sum_{a,b,i,j} H_{ab\rightarrow i j} \times  f_{a/p} \otimes f_{b/p}\otimes J_j \otimes S \times {\cal G}_i^{\psi} (E,R, z,\mu),
\eeq
where $H_{ab\rightarrow i j}$ is the hard function, $ f_{a/p}$ and $f_{b/p}$ are parton distributions functions, $J_j$ is the jet function for the jet not containing the $J/\psi$ initiated by a final state parton $j$, and ${\cal G}_i^\psi$ is the FJF for the jet containing the $J/\psi$ fragmenting from parton $i$. 
$S$ is the soft function. Generically there are two types of jets, unmeasured 
and measured, in the terminology of  Ref.~\cite{Ellis:2010rwa}. Unmeasured jet functions describe jets in which only the large light-cone momentum (measured along the jet axis) is known. 
In measured jets, some aspect of the jet's substructure has also been measured. For unmeasured jets, soft gluon radiation 
does not affect the total momentum of the jet (up to power corrections) and therefore these jet functions enter the cross section multiplicatively. For measured jets, the jet substructure may be sensitive to the soft radiation, therefore it  must be convolved with the soft function. For ${\cal G}_i^{\psi}(E,R,z, \mu)$, $R$, $E$ and $z$ are not affected by soft radiation (up to power corrections) so it also enters the cross section multiplicatively and  all of the $z$ dependence is contained in ${\cal G}_i^{\psi}(E,R,z, \mu)$, which enables us to ignore all the other factors in Eq.~(\ref{fact})
and focus on ${\cal G}_i^{\psi}(E,R,z, \mu)$.  We can therefore ignore the dependence on the other jet in Eq.~\ref{fact}, or indeed we could look at other processes with a $J/\psi$ inside a jet, such as the single-jet inclusive cross section.  In this case, there are no other jet functions and the soft function is only an overall normalization and is therefore irrelevant for our purposes.

A generic fragmenting jet function may be defined as a product of  operators of the form
\beq
{\cal G}_i^\psi = \langle 0 \mid O_{int} O_{meas} \sum_X \mid X+H\rangle \langle O+H \mid O_{int}\mid 0 \rangle,
\eeq
where $O_{int}$ is some interpolating field for the parton of interest, $i$. $O_{meas}$ is a measurement operator (a set of
delta functions) that fixes the measured jet characteristics, such as $E,R$ and $z$. The operators
are manifestly gauge invariant.
In SCET these operators would  involve only fields with the same large momentum (and possibly soft fields)
 and compose a piece of the factorization theorem (not shown) that is generated at the highest scale $Q$, which is
usually taken to be on the order of the jet energies $E$.\footnote{If there were a hierarchy then one would 
have to run these operators from the scale $Q$ to the scale $E$.}  ${\cal G}_i^\psi$ contains 
two relevant scales: the invariant mass or energy  of the jet  and the hadron mass.
Thus one can perform a further factorization to separate out these two scales where the long distance
physics is captured by a fragmentation function, and the short distance physics can be calculated
perturbatively. The resulting form of this second step of factorization can be written as
\cite{Procura:2009vm}
\beq\label{conv}
{\cal G}_i^\psi(E, R, z,\mu) = \sum_j \int_z^1 \frac{dy}{y} {\cal J}_{ij}(E, R, y, \mu) D_{j\to\psi}\left(\frac{z}{y},\mu\right)\times\left[1 + {\cal O}\left(\frac{m_\psi^2}{4E^2\tan^2(R/2)}\right)\right],
\eeq
where we have now specialized to the case of interest where the jet energy $E$ is measured for cone size $R$.\footnote{In Refs.~\cite{Procura:2009vm,Liu:2010ng,Jain:2011xz,Jain:2011iu,Procura:2011aq,Jain:2012uq} the error scales as $\Lambda_{\rm QCD}^2$ instead of $m_\psi^2$.  For our processes, the low energy scale is $m_\psi$, and thus the error scales differently.}
Loosely speaking this function gives the probability of finding a quarkonium whose  large momentum
fraction, relative to the jet within which it is found, is $z$. 
It is possible and indeed likely that there are small invariant mass jets in the data.
However, note that the process is inclusive in the sense that one integrates
over all invariant masses up to $2 E \tan[(R/2)$. So the effect of the small
invariant mass gets washed out. This is the essence of duality.
One may ask the same question about DIS, where there will be events
that contribute that are close to $x=1$, which is in the resonance region.
But if we take moments (integrating over $x$) that region gets washed out.
The differential cross section near $x=1$ is sensitive to the IR but the integrated
cross section is not.

The operator definition of the quark fragmentation function is  \cite{soper} given by
\beq
D_{j \rightarrow \psi}(z)= \frac{z}{4\pi}\int dx^+ e^{i x_+p_\psi^-/z}\frac{1}{4N_c}Tr \langle 0 \mid \bnslash \, q(x_+,0,0) \sum_X \mid X+ \psi\rangle \langle  X+\psi\mid \bar q(0) \mid 0 \rangle, 
\eeq
where the operator $q$ includes an anti-path ordered Wilson line that renders the matrix element gauge invariant.
A similar matrix element can be written down for the gluon fragmentation function.
What distinguishes the quarkonium fragmentation function from other cases is that it contains
a further subset of scales: the quark mass, the Bohr radius, and the binding energy that scale
as $1, v$, and $v^2$ respectively in units of the quark mass. Furthermore, taking the quark mass
scale to be perturbative implies that the constituents are produced at a point, and that the momentum
fraction carried by the quarkonium is set perturbatively. This is so even if the pair is produced in
an octet state, since the shedding of color occurs via soft multipole emission whose effect on
the kinematics is suppressed by an amount of order $v^2$, except near the
end point $z=1$ where these non-perturbative corrections are enhanced and can be accounted for
by the inclusion of a non-perturbative shape function \cite{BRW}. In general we will present our
results away from the end point to avoid the need for such a function.
Thus, the fragmentation functions
for quarkonium are calculable up to a set of  LDMEs.

The matching coefficients ${\cal J}_{ij}(E, R, z, \mu)$ can be calculated in perturbation theory. Large logarithms in the ${\cal J}_{ij}(E, R, z, \mu)$ are minimized at the scale
$2 E \tan(R/2)(1-z)$.  
Note that the matching coefficients   ${\cal J}_{ij}(E,R,z,\mu) $ are independent of the choice of
hadronic final states, and thus we may utilize the results 
 in Ref.~\cite{Procura:2011aq} for the FJF for light hadrons for the case at hand.

\subsection{Expressions for the $J/\psi$ FJF}
We will focus gluon and charm quark fragmentation to $J/\psi$. For gluon fragmentation to $J/\psi$ through $c\bar{c}$ pairs, we consider the $^3S_1^{(1)}$, $^3S_1^{(8)}$, $^1S_0^{(8)}$, and $^3P_J^{(8)}$ quark states. The $^3S_1^{(1)}$ gluon fragmentation function is leading order in the $v$ expansion,  as the color-octet contributions are suppressed by $v^4$. However the gluon color-singlet contribution is suppressed relative to $^3S_1^{(8)}$ by a power of  $\alpha_s^2$.  For charm quark fragmentation to $J/\psi$, we consider only  the $^3S_1^{(1)}$ contributions because both color-singlet and  color-octet mechanisms start at the same order in $\alpha_s$. The ratio of gluon to charm production cross sections  at the LHC is approximately 50, but the ratio of charm quark to gluon fragmentation functions, partially compensates for this suppression.  Fragmentation from light quarks is suppressed by one power of  $\alpha_s$ relative to the $^3S_1^{(8)}$ gluon fragmentation contribution and shares the octet velocity suppression. The ${\cal J}_{ij}(E,R,z,\mu)$ and the relevant fragmentation functions are collected in the Appendix.

The convolution in Eq.~(\ref{conv}) can be explicitly evaluated using the formula for ${\cal J}_{gg}(E,R,z,\mu)$ and ${\cal J}_{gq}(E,R,z,\mu)$ in the Appendix to obtain
\bea\label{Gg}
{\cal G}_g^\psi(E,R,z,\mu) &=& \int_z^1 \frac{d y}{y}{\cal J}_{gg}(y) D_{g\to\psi}\left(\frac{z}{y},\mu\right) + \int_z^1 \frac{d y}{y}{\cal J}_{gq}(y) D_{q\to\psi}\left(\frac{z}{y},\mu\right)   \\
&=&{\cal G}_{g(g)}^\psi(E,R,z,\mu) +{\cal G}_{g(q)}^\psi(E,R,z,\mu) \, , \nn
\eea
where 
\bea\label{Ggg}
\frac{{\cal G}_{g(g)}^\psi(E,R,z,\mu)}{2(2\pi)^3} &=& D_{g\to\psi}(z,\mu)\left(1 + \frac{C_A \alpha_s}{\pi} \left( L^2_{1-z} -\frac{\pi^2}{24}\right)\right) \\ 
&&+ \frac{C_A \alpha_s}{\pi}\left[  \int_z^1\frac{dy}{y} \tilde P_{gg}(y) L_{1-y} D_{g\to\psi}\left(\frac{z}{y},\mu \right) \right. \nn \\
&&+ 2\int_z^1dy \frac{D_{g\to\psi}(z/y,\mu)-D_{g\to\psi}(z,\mu)}{1-y } L_{1-y}  \nn \\ 
&&\left. +
\theta\left(\frac{1}{2}-z\right)
\int_z^{1/2} \frac{dy}{y} \hat{P}_{gg}(y) \ln \left(  \frac{ y}{1-y}  \right) D_{g\to\psi}\left(\frac{z}{y},\mu \right) \right] ,\nn
\eea
and 
\bea\label{Ggq}
\frac{{\cal G}_{g(q)}^\psi(E,R,z,\mu)}{2(2\pi)^3} &=& \frac{T_F \alpha_s}{\pi}\left[  \int_z^1\frac{dy}{y} \left[ P_{qg}(y) L_{1-y} + y(1-y)\right] D_{q\to\psi}\left(\frac{z}{y},\mu \right) \right. \\
&& \left. +
\theta\left(\frac{1}{2}-z\right)
\int_z^{1/2} \frac{dy}{y} {P}_{qg}(y) \ln \left(  \frac{ y}{1-y}  \right) D_{q\to\psi}\left(\frac{z}{y},\mu \right) \right] \nn \,.
\eea
In this expression, we have defined
\bea
L_{1-z} &=& \ln \left(\frac{2 E \tan(R/2)(1-z)}{\mu}\right), \nn \\
\hat P_{gg}(z) &=& 2\left[\frac{z}{(1-z)_+}+\frac{1-z}z+ z(1-z)\right] ,\nn\\
\tilde P_{gg}(z) &=& 2\left[\frac{1-z}z+ z(1-z)\right] ,\nn\\
P_{qg}(z) &=&  z^2+(1-z)^2 \nn\,.
 \eea
This expression  shows that the logarithms in ${\cal G}^\psi_g(E,R,z,\mu)$ are minimized at the scale $\mu = 2 E \tan(R/2)(1-z)$, as first pointed out in Ref.~\cite{Procura:2011aq}.
  The logarithms of $1-z$ are easily resummed using the jet anomalous dimension~\cite{Procura:2011aq},
however, we will not do this resummation  in this paper as we consider $1-z \sim {\cal O}(1)$. We instead set the scale in ${\cal J}_{gg}(E,R,z,\mu)$ to be $\mu_J = 2E \tan(R/2)$, and evolve the fragmentation function from the scale $2 m_c$  to the scale  $\mu_J$.  This is done by taking moments of the fragmentation functions, evolving each moment according to its anomalous dimension as obtained from the Altarelli-Parisi equations, and then performing an inverse-Mellin transform. 

The ${\cal G}_{g(q)}^\psi(E,R,z,\mu)$ is present because of mixing with the quark fragmentation function. In principle there should be a sum over all quark flavors. However, the light quark fragmentation function 
contributes only via fragmentation through $^3S_1^{(8)}$ $c\bar{c}$ pairs at $O(\alpha_s^2)$ and is subleading to the $^3S_1^{(8)}$ gluon fragmentation so it will be neglected. 
Charm quarks and antiquarks  can fragment via   $^3S_1^{(1)}$ $c\bar{c}$ pairs at $O(\alpha_s^2)$, which is lower order than the corresponding gluon fragmentation function. 
Therefore this mixing must be included.

The quark FJF is given by:
\bea\label{Gq}
\frac{{\cal G}_{q}^\psi(E,R,z,\mu)}{2(2\pi)^3} &=& D_{q\to\psi}(z,\mu)\left(1 + \frac{C_F \alpha_s}{\pi} \left( L^2_{1-z} -\frac{\pi^2}{24}\right)\right)  \\
&& + \frac{C_F \alpha_s}{\pi}\left[  \int_z^1\frac{dy}{y} (1-y)\left(L_{1-y}+\frac{1}{2}\right) D_{q\to\psi}\left(\frac{z}{y},\mu \right) \right. \nn \\
&& + 2\int_z^1dy \frac{D_{q\to\psi}(z/y,\mu)-D_{q\to\psi}(z,\mu)}{1-y } L_{1-y}  \nn \\
&& \left. +
\theta\left(\frac{1}{2}-z\right)
\int_z^{1/2} \frac{dy}{y} \hat{P}_{qq}(y) \ln \left(  \frac{ y}{1-y}  \right) D_{q\to\psi}\left(\frac{z}{y},\mu \right) \right] \nn \\
&&+ \frac{C_F\alpha_s }{\pi} \left[ \int_z^1 \frac{dy}{y}\left( P_{gq}(z) L_{1-y} +\frac{y}{2} \right) D_{g\to \psi}\left(\frac{z}{y},\mu\right) \nn \right. \\
&&\left. +\theta\left(\frac{1}{2}-z\right) \int_z^{1/2} \frac{dy}{y} \log\left(\frac{y}{1-y}\right) P_{gq}(y) D_{g\to \psi}\left(\frac{z}{y},\mu\right) \right] \nn
\, ,
\eea
where
\bea 
\hat P_{qq}(z) &=& \frac{1+z^2}{(1-z)_+} ,\nn \\
P_{gq}(z) &=& \frac{1+(1-z)^2}{z} \, . \nn
\eea
For this contribution, as previously mentioned we will only consider the $q=c$ contribution fragmenting via $^3S_1^{(1)}$ $c\bar{c}$ pairs. The mixing contribution of gluon fragmentation into this FJF must also be included.
To evaluate ${\cal G}^{\psi}_i(E,R,z,\mu_J)$ we will use Eqs.~(\ref{Gg}-\ref{Gq}) with our numerically evaluated $D_{i\to \psi}(z,\mu_J)$.
We see that up to $O(\alpha_s)$ corrections
\bea\label{approx}
\frac{{\cal G}_i^\psi(E,R,z,\mu_J)}{2(2\pi)^3} \to  D_{i\to\psi}(z,\mu_J) + O(\alpha_s(\mu_J)) ,
\eea
which shows that the $z$ distribution of a $J/\psi$ within a jet with energy $E$ and cone size $R$ is approximately equal to the fragmentation function evaluated 
at  the jet  scale $\mu_J =2E \tan(R/2)$.  

Since the fragmentation functions for  $^3S_1^{(1)}$, $^3S_1^{(8)}$, $^1S_0^{(8)}$, and $^3P_J^{(8)}$ are very different, this observable has the power to discriminate between all four gluon-production mechanisms. This can seen from a cursory inspection of the expressions for the fragmentation functions given in the Appendix and shown in Fig.~\ref{ffschem}. 
\begin{figure}[!t]
\begin{center}
\includegraphics[width=8cm]{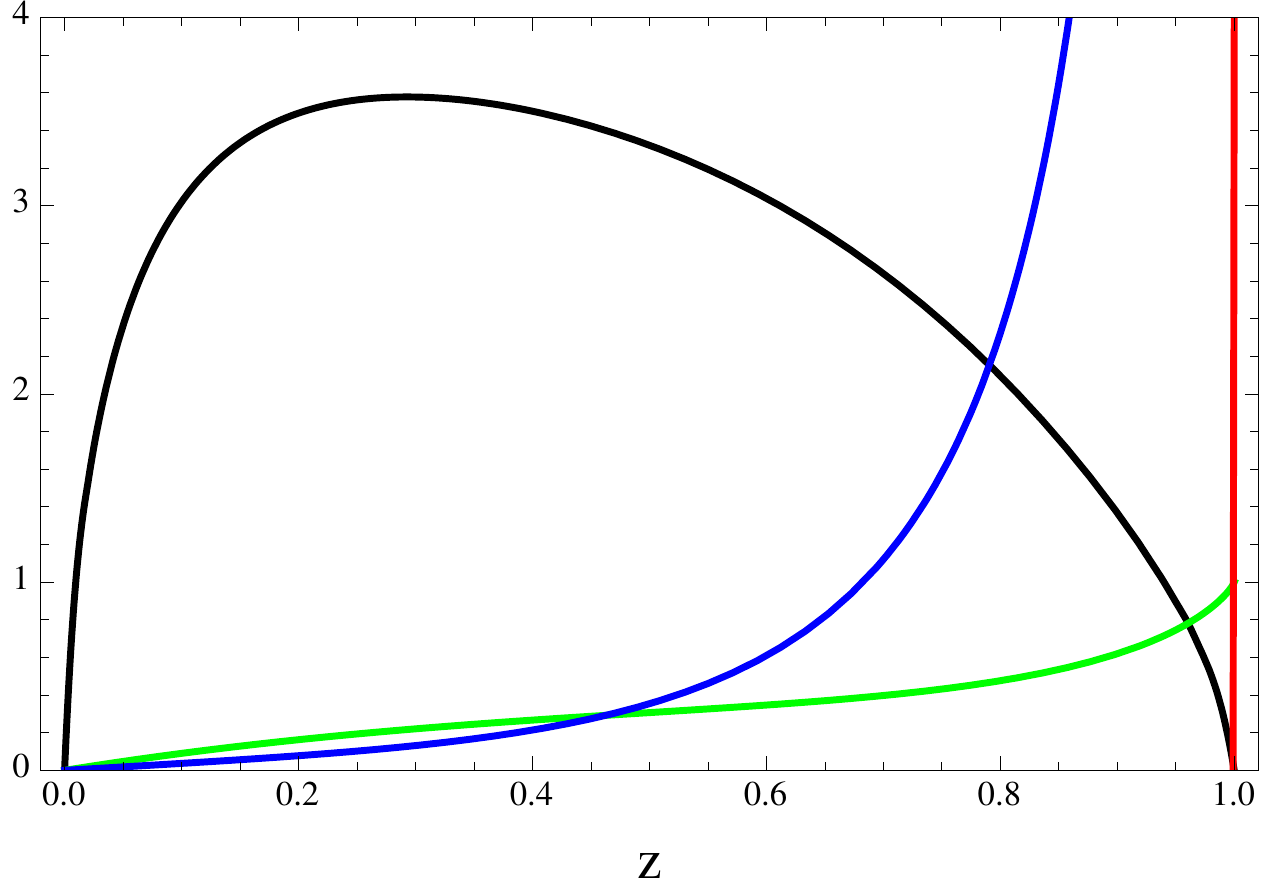}
\end{center}
\vspace{- 0.5 cm}
\caption{
\label{ffschem} 
\baselineskip 3.0ex
The gluon fragmentation functions at $\mu = 2m_c$ for $^3S_1^{(1)}$ (black), $^3S_1^{(8)}$ (red), $^1S_0^{(8)}$ (green), $^3P_J^{(8)}$ (blue).  Relative normalization is arbitrary and relevant formulas are found in the Appendix.}
\end{figure}
Though the dramatic differences in these functions are considerably softened by Altarelli-Parisi evolution, we will see  that each contribution to ${\cal G}_g^\psi(E,R,z,\mu)$ has a different $E$ dependence that varies for fixed $z$ ({\it cf.}~Fig.~\ref{fixedz}).
 This makes it clear that measurement of ${\cal G}_g^\psi(E,R,z,\mu)$ for different momentum fractions has potential to allow independent extraction of all four LDME. In our  calculations $E$ and $R$ will always enter in the combination $E\tan(R/2)$ and we will choose $R=0.4$ .

\begin{figure}[!t]
\begin{center}
\includegraphics[width=16cm]{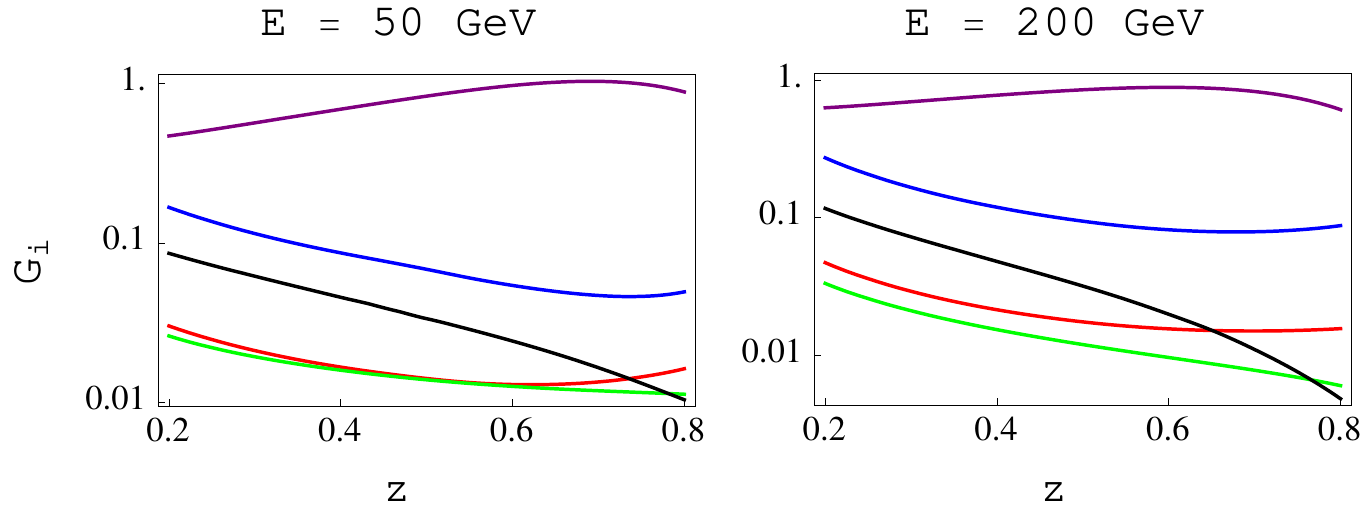}
\end{center}
\vspace{- 0.5 cm}
\caption{
\label{logFJF} 
\baselineskip 3.0ex
The gluon FJF (color coding the same as in Fig.~\ref{ffschem}) and the charm quark FJF for $^3S_1^{(1)}$ (purple).   }
\end{figure}

In Fig.~\ref{logFJF} we plot the $^3S_1^{(1)}$ (black), $^3S_1^{(8)}$ (red), $^1S_0^{(8)}$ (green), and $^3P_J^{(8)}$ (blue) gluon FJFs as well  as the $^3S_1^{(1)}$ charm (purple) FJF for $E = 50$ GeV and $E= 200$ GeV. This plot illustrates the discriminating power of the jet observables. For  Fig.~\ref{logFJF} we have chosen the LDME to be the central values extracted in the fits of Refs.~{\cite{Butenschoen:2011yh,Butenschoen:2012qr}: $\langle {\cal O}^{J/\psi}(^3S_1^{(1)})\rangle = 1.32 \, {\rm GeV}^3$,  
$\langle  {\cal O}^{J/\psi}(^1S_0^{(8)})\rangle = 4.97 \times 10^{-2} \,{\rm GeV}^3$, 
$\langle  {\cal O}^{J/\psi}(^3S_1^{(8)})\rangle = 2.24 \times 10^{-3} \,{\rm GeV}^3$,  and
$\langle  {\cal O}^{J/\psi}(^3P_0^{(8)})\rangle = -1.61 \times 10^{-2} \,{\rm GeV}^5$.  Throughout this work we take $m_c$ = 1.4 GeV.

It is also interesting to study the energy dependence of the fragmentation functions. In Fig.~\ref{fixedz} we plot the four gluon FJFs 
as a function of energy $E$ for three different values of $z$ using the same color-coding as above. The LDME of Refs.~\cite{Butenschoen:2011yh,Butenschoen:2012qr} 
have again been used to set the normalization of the  curves. In order to the make shapes of the curves more easily viewable, we 
have divided the $^3P_J^{(8)}$ by a factor of 5 and the color-singlet contribution has been divided by a factor of 2. The shapes 
of the energy dependence at different values of $z$ are quite distinct for all four fragmentation functions. For example, the $^3P_J^{(8)}$ 
FJF is an increasing function of energy for all three $z$ values, while the $^1S_0$ and the color-singlet are decreasing  functions 
of $E$ for $z$ = 0.5 and 0.8, and the $^3S_1^{(8)}$ is decreasing only for 0.8. Extractions of the  $E$ dependence of the FJF for
different values of $z$ should allow one to disentangle the various contributions to quarkonium production.
In particular, note that if the lack of polarization is due to an anomalously large ${}^1S_0^{(8)}$,
then we should see a decrease in the gluon FJF as a function of the jet energy for fixed $z$, with $z \, >$ 0.5.
\begin{figure}[!t]
\begin{center}
\includegraphics[width=16cm]{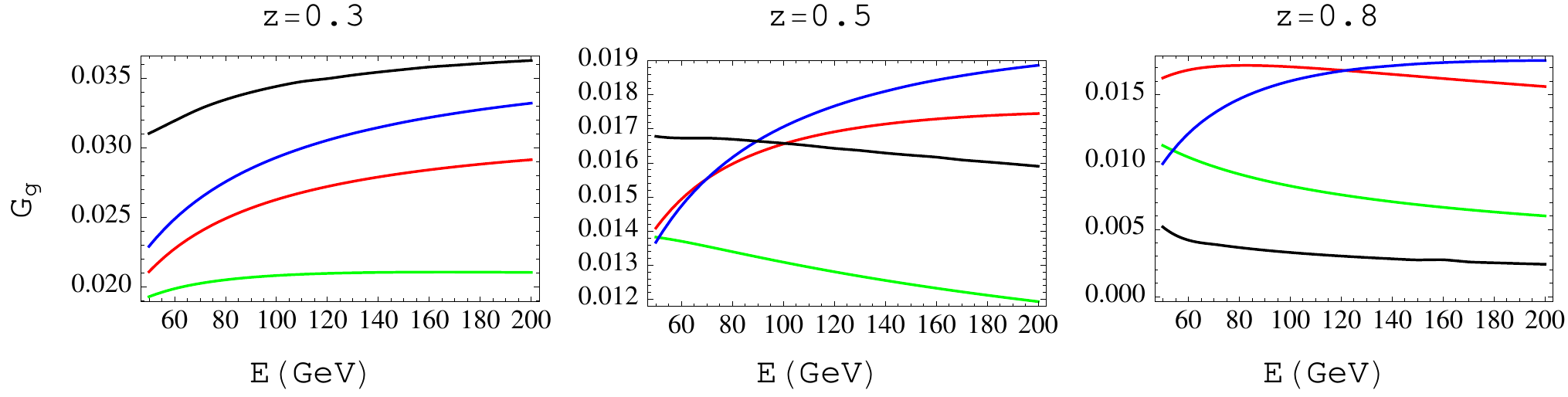}
\end{center}
\vspace{- 0.5 cm}
\caption{
\label{fixedz} 
\baselineskip 3.0ex
The energy dependence of the four different contributions to the  gluon FJF for fixed $z=0.3$, $0.5$, and $0.8$. Color coding is the same as in Figs.~\ref{ffschem}, \ref{momfig}.  For readability, we have scaled the $^3P_J^{(8)}$ function down by a factor of 5 and $^3S_1^{(1)}$ down by 2.  These plots have been normalized with respect to the total rate and thus do not reflect its underlying energy dependence.}
\end{figure}
\begin{figure}[!t]
\begin{center}
\includegraphics[width=16cm]{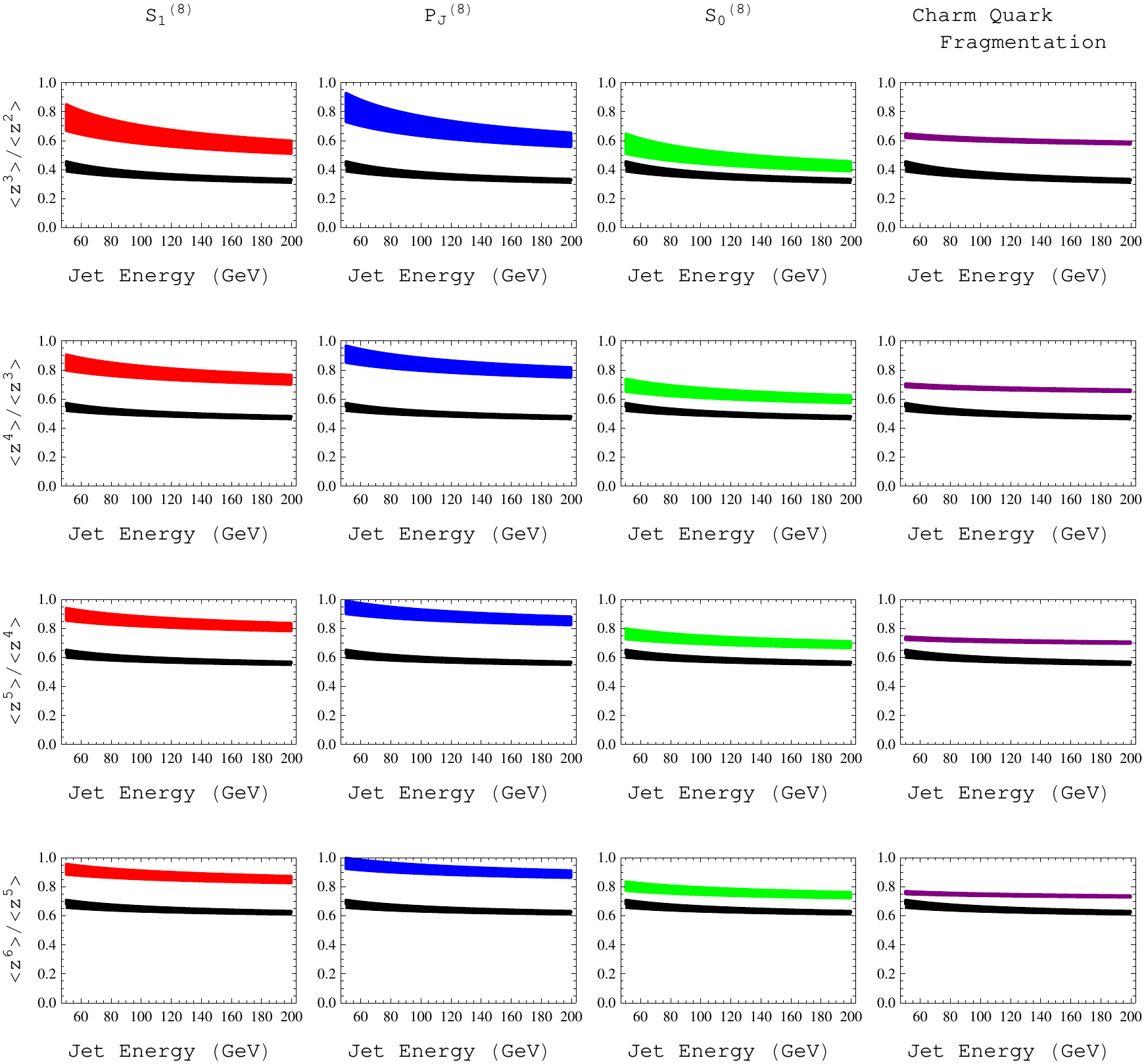}
\end{center}
\vspace{- 0.5 cm}
\caption{
\label{momfig} 
\baselineskip 3.0ex
Ratios of successive moments as a function of the jet energy. See text for explanation.}
\end{figure}
The moments of the FJF, $\langle z^N\rangle \equiv \int_0^1 dz\, z^{N-1} {\cal G}_g^\psi(E,R,z,\mu)$, can be calculated analytically using the formulae in the Appendix. Note that this 
integral diverges if $N=1$ because  the $N=1$ moments of both the Altarelli-Parisi splitting function and the matching coefficients ${\cal J}_{gg}(E,R,z,\mu)$
have poles at $N=1$. This could be cured by resummation of $\log z$, as implemented for the $D_{g\to\psi}(z,\mu)$  fragmentation function in Ref.~\cite{Boyd:1998km}, but 
this is beyond the scope of this paper. The LDME cancel in the ratios of moments, and we plot ratios of successive moments,  $\langle z^{N+1}\rangle/ \langle z^N\rangle$, for $N=2,3,$ and $4$
in Fig.~\ref{momfig}. In all columns we have plotted the moment ratios of the $^3S_1^{(1)}$ FJF (black).
We also plot moment ratios for the 
 $^3S_1^{(8)}$ FJF (red),  $^3P_J^{(8)}$  (blue),  $^1S_0^{(8)}$ FJF (green), and the charm quark FJF (purple), in each column respectively. Scale uncertainties are included  by varying $E \tan(R/2) < \mu < 4E \tan(R/2)$. We see that the moments have power to discriminate between various production mechanisms, in particular, we find 
\bea
\frac{\langle z^{n+1}\rangle}{\langle z^{n}\rangle }\bigg|_{^3P_J^{(8)}} \approx \frac{\langle z^{n+1}\rangle}{\langle z^{n}\rangle }\bigg|_{^3S_1^{(8)}} > 
\frac{\langle z^{n+1}\rangle}{\langle z^{n}\rangle }\bigg|_{^1S_0^{(8)}} \approx \frac{\langle z^{n+1}\rangle}{\langle z^{n}\rangle }\bigg|_{\rm c-quark}
>  \frac{\langle z^{n+1}\rangle}{\langle z^{n}\rangle }\bigg|_{^3S_1^{(1)}} \, .
\eea
Note that for the same choice of $\mu$,  
\bea
\frac{\langle z^{n+1}\rangle}{\langle z^{n}\rangle }\bigg|_{^3P_J^{(8)}} > \frac{\langle z^{n+1}\rangle}{\langle z^{n}\rangle }\bigg|_{^3S_1^{(8)}} \, ,
\eea 
but once scale uncertainties are included it is hard to distinguish these two moment ratios. The energy dependence of the moments of the color-octet FJFs is given by
\bea\label{momeq}
\langle z^N \rangle =\tilde{ {\cal J}}_{gg}(E,R,N,\mu) \left(\frac{\alpha_s(\mu)}{\alpha_s(2m_c)}\right)^{2\gamma^N_{gg}/ b_0} \tilde{D}_{g\to\psi}(N,2 m_c) \, .
\eea  
When we set $\mu \approx 2 E \tan(R/2)$, the energy dependence is entirely contained in the first two factors on the r.h.s.~of Eq.~(\ref{momeq}), which are the same for all three
color-octet FJFs. The color-singlet and charm quark fragmentation functions are more complicated due to the mixing of these fragmentation 
function in the evolution from the scale $2 m_c$ to $\mu_J$. Making log-log plots of $\langle z^N\rangle$ we find that that  $\langle z^N\rangle \propto (\log E)^{F(N)}$ where $F(N)$ can be extracted from Eq.~(\ref{momeq}).

\section{Comparison of Various LDME Extractions}

In the final part of this paper, we will discuss what recent extractions of the LDME predict for the gluon FJF. In addition to the extractions in Refs.~\cite{Butenschoen:2011yh,Butenschoen:2012qr}, we will consider values of the LDME extracted in two recent papers~\cite{Chao:2012iv,Bodwin:2014gia} that attempt to solve the polarization puzzle by focusing exclusively on high $p_\perp$ production of charmonia at collider experiments. 
The study in Ref.~\cite{Chao:2012iv} uses a NLO NRQCD calculation to fit the  color-octet  LDME  to inclusive $J/\psi$ production at high $p_\perp$ and finds 
values of the LDME that can produce negligible polarization in agreement with the data. However, these values of LDME are inconsistent with the results of fitting the world data in Refs.~\cite{Butenschoen:2011yh,Butenschoen:2012qr}. In particular, $\langle {\cal O}^{J/\psi}(^1S_0^{(8)})\rangle$ is larger by a factor of two and $\langle {\cal O}^{J/\psi}(^3P_0^{(8)})\rangle $ has the opposite sign as the fit
 in Refs.~\cite{Butenschoen:2011yh,Butenschoen:2012qr}.  These two effects combine to produce significant depolarization of the $J/\psi$. In Ref.~\cite{Bodwin:2014gia}, the calculations are performed in the leading-power fragmentation approximation and logarithms of $p_\perp/m_c$ are resummed by using Altarelli-Parisi equations for the fragmentation functions. 
 The fitted LDME are similar to those found in Ref.~\cite{Chao:2012iv} in the sense that  $\langle {\cal O}^{J/\psi}(^1S_0^{(8)})\rangle$ is by far the largest matrix element and $\langle {\cal O}^{J/\psi}(^3P_0^{(8)}) \rangle $  again has opposite sign as that extracted from fits to the world data. In Ref.~\cite{Bodwin:2014gia},  the errors on $\langle {\cal O}^{J/\psi}(^3S_1^{(8)})\rangle$ and $\langle {\cal O}^{J/\psi}(^3P_0^{(8)}) \rangle$ are  essentially 100\% so the extracted matrix elements are consistent with zero. This analysis suggests that the production of $J/\psi$ at large $p_\perp$  is dominated by $c\bar{c}$ pairs in a $^1S_0^{(8)}$ state rather than  $^3S_1^{(8)}$. It should be noted that the quoted errors in the extracted LDME in Refs.~\cite{Butenschoen:2011yh,Butenschoen:2012qr} are considerably smaller than those in Refs.~\cite{Chao:2012iv,Bodwin:2014gia}. However, the presence of nontrivial correlations between the uncertainties in \cite{Bodwin:2014gia} allows us to make a much sharper prediction for the gluon FJF than is naively suggested by the large individual error bars \cite{Bodwin:2014email}.  In all of these extractions, there is a hierarchy between matrix elements that are supposed to have the same velocity scaling. However, it is generated by anomalously small matrix elements not anomalously large ones.

\begin{figure}[!t]
\begin{center}
\includegraphics[width=16cm]{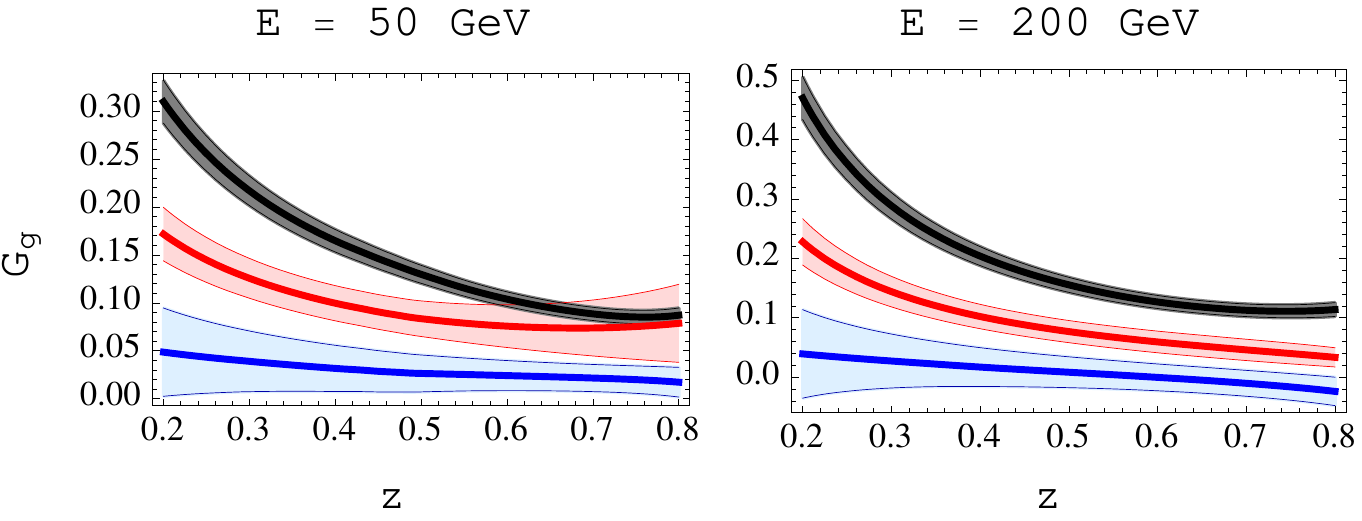}
\end{center}
\vspace{- 0.5 cm}
\caption{
\label{fitcompare} 
\baselineskip 3.0ex
The gluon FJF at fixed energy for the LDME extracted in Refs.~\cite{Butenschoen:2011yh,Butenschoen:2012qr} (gray),
Ref.~\cite{Chao:2012iv} (blue), and Ref.~\cite{Bodwin:2014gia} (red). }
\end{figure}

\begin{figure}[!t]
\begin{center}
\includegraphics[width=16cm]{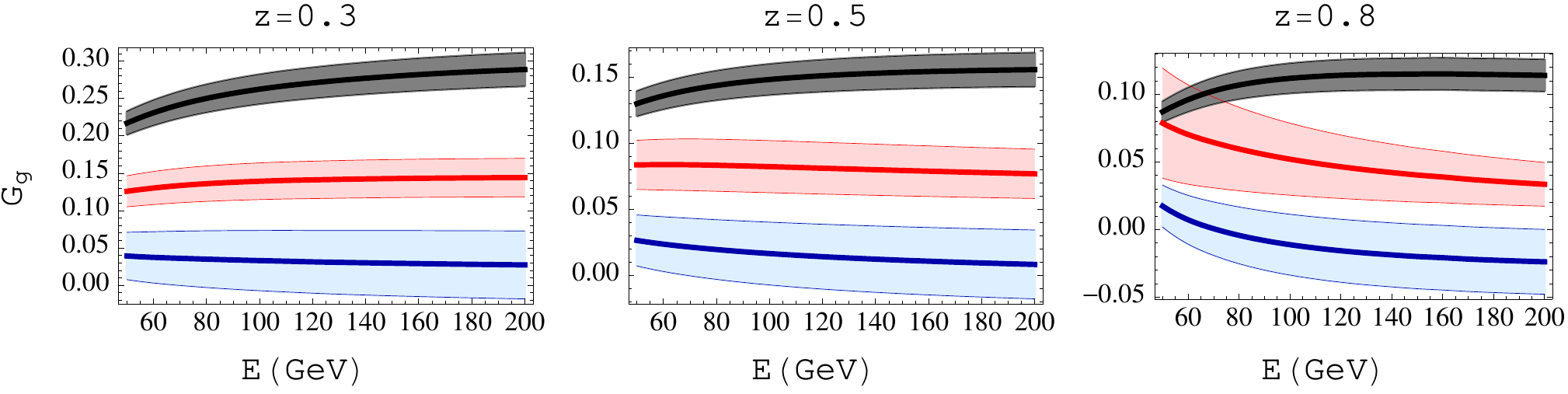}
\end{center}
\vspace{- 0.5 cm}
\caption{
\label{zcompare} 
\baselineskip 3.0ex
The gluon FJF at fixed momentum fraction for the LDME extracted in Refs.~\cite{Butenschoen:2011yh,Butenschoen:2012qr} (gray),
Ref.~\cite{Chao:2012iv} (blue), and Ref.~\cite{Bodwin:2014gia} (red). These plots have been normalized with respect to the total rate.}
\end{figure}

In Fig.~\ref{fitcompare}, we compare the predictions for the gluon FJF at $E=50$ GeV and $E = 200$ GeV using the results from the fits to the LDME 
in Refs.~\cite{Butenschoen:2011yh,Butenschoen:2012qr, Chao:2012iv,Bodwin:2014gia}. The gluon FJF is the sum over all contributions,
color-singlet as well as  color-octet. The color-singlet matrix element is chosen to be 1.32 ${\rm GeV}^3$
in Refs.~\cite{Butenschoen:2011yh,Butenschoen:2012qr,Bodwin:2014gia} and 1.16 ${\rm GeV}^3$ in Ref.~\cite{Chao:2012iv}.
We use the LDME extracted in the original fit and the error bands are the result of adding in quadrature the uncertainties for the LDME quoted in Refs.~\cite{Butenschoen:2011yh,Butenschoen:2012qr, Chao:2012iv}. We supplement the uncertainty given in Ref.~\cite{Bodwin:2014gia} with the full correlation matrix provided by one of the authors \cite{Bodwin:2014email}. No other theoretical uncertainty is included. The gray band with black borders is the prediction using the LDME extracted in Refs.~\cite{Butenschoen:2011yh,Butenschoen:2012qr}, the red band uses the matrix elements extracted in Ref.~\cite{Bodwin:2014gia}
and the blue band uses the matrix elements extracted in Ref.~\cite{Chao:2012iv}. Fig.~\ref{zcompare} shows the energy dependence at fixed momentum fraction for the different determinations.  We see that for $z >$0.5, the question of which set of LDMEs is preferred, those determined for the world average \cite{Butenschoen:2011yh,Butenschoen:2012qr} or those that alleviate the polarization puzzle \cite{Chao:2012iv,Bodwin:2014gia}, will be resolved by testing whether the gluon FJF is increasing or decreasing with energy.  Furthermore, measurement of the gluon FJF has the power to distinguish between all three fits.

\section{Conclusions}
We have demonstrated that by studying the characteristics of jets arising from
quarkonium production, we can disentangle the various production channels.
There are a multitude of ways of analyzing such events. Here we have chosen
to measure the energy and cone angle of the jet, but one could consider
other observables such as the invariant mass.
Within our choice of variables ($E,R$) we found that a particularly discriminating tool is the measurement of the
energy dependence at fixed momentum fraction as shown in Figs.~\ref{fixedz} and \ref{zcompare}.
A robust prediction of our analysis is that for $z\, >$ 0.5 the gluon FJF at fixed $z$ should decrease as function
of energy if the lack of transverse polarization in the data is due to the dominance of the ${}^1S_0^{(8)}$ LDME over the other color octet matrix elements for high-$p_\perp$ production.  Further information can be gathered by calculating the normalized cross section, 
in which case one could constrain the sum of the matrix elements. 

\acknowledgments 

TM and AKL acknowledge support from  the ESI workshop, Jets in Quantum Field Theory, where this work was initiated. MB acknowledges the Center for Future High Energy Physics at IHEP where a portion of this work was completed.
We  thank Andrew Hornig, Wouter Waalewijn, Massimiliano Procura, Geoffrey Bodwin, and Christian Bauer for useful discussions and James Russ for comments on the manuscript.  AKL was supported in part by the National Science Foundation under Grant No. PHY-1212635. TM was supported in part by the Director, Office of Science, Office of Nuclear Physics, of the U.S. Department of Energy under grant numbers DE-FG02-05ER41368.  IZR and MB  are supported by DOE DE-FG02-04ER41338 and FG02-06ER41449.

\appendix
\section{Formulae For Matching Coefficients, Fragmentation Functions,  Moments}

In this appendix we collect the basic formulae needed for the calculation.  The matching coefficients ${\cal J}_{ij}(E,R,z,\mu)$ are calculated in Ref.~\cite{Procura:2011aq}:\bea
\frac{{\cal J}_{gg}(E, R, z, \mu)}{2(2\pi)^3} &=& \delta (1-z) + \frac{\alpha_s(\mu)C_A} \pi \left[
\left( L^2-\frac{\pi^2}{24} \right) \delta(1-z) + \hat P_{gg}(z)L + \hat {\cal J}_{gg}(z)
\right], \\
\frac{{\cal J}_{qq}(E, R, z, \mu)}{2(2\pi)^3} &=& \delta (1-z) + \frac{\alpha_s(\mu)C_F} \pi \left[
\left( L^2-\frac{\pi^2}{24} \right) \delta(1-z) + \hat P_{qq}(z)L + \hat {\cal J}_{qq}(z)
\right],   \\
\frac{{\cal J}_{gq}(E, R, z, \mu)}{2(2\pi)^3} &=& \frac{\alpha_s(\mu)T_F} \pi \left[
 P_{qg}(z)L + \hat {\cal J}_{gq}(z)   
\right],   \\
\frac{{\cal J}_{qg}(E, R, z, \mu)}{2(2\pi)^3} &=& \frac{\alpha_s(\mu)C_F} \pi \left[
 P_{gq}(z)L + \hat {\cal J}_{qg}(z)   
\right]   ,  
\eea
where $L = \ln[2 E \tan(R/2)/\mu]$, and
\bea
\hat {\cal J}_{gg}(z) &=& \left\{
\begin{array}{lc}
\hat P_{gg}(z)\ln z& z \leq 1/2\\
\frac{2(1-z+z^2)^2}z \left(\frac{\ln(1-z)}{1-z}\right)_+\qquad& z \geq 1/2
\end{array} 
\right.  , \\
\hat {\cal J}_{qq}(z) &=& \frac{1}{2}(1-z) + \left\{
\begin{array}{lc}
\hat P_{qq}(z)\ln z& z \leq 1/2\\
(1+z^2) \left(\frac{\ln(1-z)}{1-z}\right)_+\qquad& z \geq 1/2
\end{array} 
\right.  , \\
\hat {\cal J}_{gq}(z) &=& z(1-z) + P_{qg}(z) \left\{
\begin{array}{lc}
\ln z& z \leq 1/2\\
\ln(1-z)\qquad& z \geq 1/2
\end{array} 
\right. ,\\
\hat {\cal J}_{qg}(z) &=& \frac{z}{2}  + P_{gq}(z) \left\{
\begin{array}{lc}
\ln z& z \leq 1/2\\
\ln(1-z)\qquad& z \geq 1/2
\end{array} 
\right. \, .
\eea

There are five NRQCD  fragmentation functions.  The $^3S_1^{(8)}$ gluon fragmentation function  is given by~\cite{Braaten:1994vv}  
\beq
D_{g\to \psi}^{^3S_1^{(8)}}(z,2m_c)  = \frac{\pi\alpha_s(2m_c)}{24 m_c^3}\langle O^\psi(^3S_1^{(8)})\rangle \delta(1-z),
\eeq
and the  $^3S_1^{(1)}$ gluon fragmentation function is  \cite{Braaten:1993rw,Ma:2013yla}
\bea
D_{g\to \psi}^{^3S_1^{(1)}}(z,2m_c) &=& \frac{5\alpha_s^3(2m_c)}{864\pi}\frac{\langle O^\psi(^3S_1^{(1)})\rangle}{m_c^3}\int_0^z dr\int_{(r+z^2)/2z}^{(1+r)/2} dy\frac1{(1-y)^2(y-r)^2(y^2-r)^2}\nn \\
&&\sum_{i=0}^2 z^i\left(f_i(r,y) + g_i(r,y)\frac{1+r-2 y}{2(y-r)\sqrt{y^2-r}}\ln\frac{y-r+\sqrt{y^2-r}}{y-r-\sqrt{y^2-r}}\right)\!,
\eea
where
\bea
f_0(r,y) &=& r^2(1+r)(3+12r+13r^2) - 16 r^2(1+r)(1+3r)y\nn\\
&& - 2 r(3 - 9 r - 21 r^2 + 7 r^3)y^2 + 8 r(4 + 3 r + 3 r^2) y^3 - 4 r(9-3r-4r^2) y^4\nn\\
&&-16(1+3r+3r^2)y^5 + 8(6+7r)y^6-32 y^7, \nn \\
f_1(r,y) &=& - 2 r(1+5r+19r^2 + 7r^3)y + 96r^2(1+r)y^2 +8(1-5r-22r^2-2r^3)y^3\nn\\
&&+16 r(7+3r)y^4 - 8(5+7r)y^5+32 y^6,\nn \\
f_2(r,y) &=& r(1+5r+19r^2+7r^3) - 48r^2(1+r)y - 4(1-5r-22 r^2-2r^3)y^2\nn\\
&& - 8 r(7+3r)y^3 + 4(5+7r)y^4 - 16 y^5,\nn \\
g_0(r,y) &=& r^3(1-r)(3 + 24 r+ 13 r^2) - 4r^3(7-3r-12 r^2)y - 2 r^3(17+22r-7r^2)y^2\nn\\
&&+ 4r^2(13+5r - 6r^2)y^3 - 8 r(1+2r+5r^2+2r^3)y^4 - 8 r(3-11r-6r^2)y^5\nn\\
&&+8(1-2r-5r^2)y^6,\nn \\
g_1(r,y) &=& - 2 r^2(1+r)(1-r)(1+7r)y + 8 r^2(1+3r)(1-4r)y^2\nn\\
&& + 4 r(1+10 r + 57 r^2 + 4 r^3)y^3 - 8 r(1+29 r + 6 r^2)y^4 - 8(1-8r-5r^2)y^5,\nn \\
g_2(r,y) &=& r^2(1+r)(1-r)(1+7r) - 4r^2(1+3r)(1-4r)y\nn\\
&&-2r(1+10r+57r^2+4r^3)y^2 + 4 r(1+29 r + 6 r^2)y^3 + 4(1-8r-5r^2)y^4. \nn
\eea
The integrals over $r$ and $y$ must be done numerically. The $^1S_0^{(8)}$ gluon fragmentation function  is given by~\cite{Braaten:1996rp,Bodwin:2012xc,Ma:2013yla}
\bea
D_{g\to \psi}^{^1S_0^{(8)}}(z,2m_c)  = \frac{5 \alpha_s^2(2 m_c)}{96 m_c^3}  \langle {\cal O}^\psi(^1S_0^{(8)})\rangle \left(3 z -2z^2+2(1-z)\log(1-z)\right) \, ,
\eea
and the  $^3P_J^{(8)}$ gluon fragmentation function is given by
\bea
D_{g\to \psi}^{^3P_J^{(8)}}(z,2m_c) &=&  \frac{5 \alpha_s^2(2 m_c)}{12 m_c^5}  \langle {\cal O}^\psi(^3P_0^{(8)})\rangle  \\
&&\times\left(\frac{1}{6}\delta(1-z) + \frac{1}{(1-z)_+}+ \frac{13-7z}{4}\log(1-z)-\frac{(1-2z)(8-5 z)}{8}\right) \, .\nn
\eea
Here we have summed over $J=0,1,2$ and used $ \langle {\cal O}^\psi(^3P_J^{(8)})\rangle =(2J+1) \langle {\cal O}^\psi(^3P_0^{(8)})\rangle$.
The  $^3S_1^{(1)}$ charm quark  fragmentation function  is  \cite{Braaten:1993rw}, 
\bea
\!\! D_{c\to \psi}^{^3S_1^{(1)}}(z,2m_c) &=& \frac{32\alpha_s^2(2m_c)}{81}\frac{\langle O^\psi(^3S_1^{(1)})\rangle}{m_c^3} 
\frac{(z-1)^2}{(z-2)^6}\, z(5z^4 -32 z^3+72 z^2-32 z +16 ) \, .
\eea

The moments of the color-octet gluon  fragmentation functions can be computed analytically. Defining
\bea
\tilde{D}_{g\to \psi}(N,2m_c) = \int_0^1dz z^{N-1} D_{g\to \psi}(z,2m_c) \, ,
\eea 
we have 
\bea
\tilde{D}_{g\to \psi}^{^3S_1^{(8)}}(N,2m_c)  &=& \frac{\pi\alpha_s(2m_c)}{24 m_c^3}\langle O^\psi(^3S_1^{(8)})\rangle, \\
\tilde{D}_{g\to \psi}^{^1S_0^{(8)}}(N,2m_c) &=& \frac{5 \alpha_s^2(2 m_c)}{96 m_c^3}  \langle {\cal O}^\psi(^1S_0^{(8)})\rangle
\left[\frac{8+7 N+N^2}{(N+1)^2(N+2)}- \frac{2 H_N}{N(N+1)}\right] ,  \\
\tilde{D}_{g\to \psi}^{^3P_J^{(8)}}(N,2m_c) &=& \frac{5 \alpha_s^2(2 m_c)}{12 m_c^5}  \langle {\cal O}^\psi(^3P_0^{(8)})\rangle \\
&& \times\left[\frac{188+191 N+49N^2+4N^3}{24(N+1)^2(N+2)}-\frac{4N^2+10 N+13}{4N(N+1)}H_N\right] \, .  \nn
\eea

The fragmentation function is evolved using the standard DGLAP evolution,
\beq
\mu\frac{\partial}{\partial \mu} D_i(z,\mu) = \frac{\alpha_s(\mu)}\pi \sum_j\int_z^1\frac{dy}y P_{i\to j}(z/y,\mu) D_j(y,\mu) \, .
\eeq
These equations are solved analytically in moment space, and then the fragmentation functions at the scale $\mu_J$ are obtained by numerically evaluating the inverse Mellin transform.  In our calculations, $q=c$ and the mixing between the gluon and $c$ quark fragmentation function 
is only relevant for the $^3S_1^{(1)}$ channel.

It is useful to have analytic expressions for the moments of the matching coefficients; these are given by:
\bea
\tilde{{\mathcal J}}_{gg}(E,R,N,\mu) &=& \int_0^1 dz \, z^{N-1} \frac{{\cal J}_{gg}(E,R,z,\mu)}{2(2\pi)^3} \nn \\
&=&1 + \frac{\alpha_s C_A}{\pi} \left( L^2 + P_{gg}^N L + H_{N-1}^2  - \frac{5\pi^2}{24} + H_{N-1,2} \right. \nn \\
&&\quad \!  + \,2\, G_{N-1}+F_{N-2} -2 F_{N-1} +F_{N}-F_{N+1}  \bigg) , \\
\tilde{{\mathcal J}}_{qq}(E,R,N,\mu) &=& \int_0^1 dz \, z^{N-1} \frac{{\cal J}_{qq}(E,R,z,\mu)}{2(2\pi)^3}  \nn \\
&=&1 + \frac{\alpha_s C_F}{\pi} \left( L^2 + P_{qq}^N L + \frac{H_{N-1}^2 + H_{N+1}^2 }{2}   - \frac{5\pi^2}{24} + 
\frac{H_{N-1,2}+H_{N+1,2}}{2}  \right. \nn \\
&&\quad \!  + G_{N-1}  + G_{N+1}  \bigg)  , \\
\tilde{{\mathcal J}}_{gq}(E,R,N,\mu) &=& \int_0^1 dz \, z^{N-1} \frac{{\cal J}_{gq}(E,R,z,\mu)}{2(2\pi)^3} \nn \\
&=& \frac{\alpha_s T_F}{\pi} \left(   P_{qg}^N L + \frac{1}{(N+1)(N+2)}  + F_{N+1}-F_N+\frac{1}{2}F_{N-1} \right), \\
\tilde{{\mathcal J}}_{qg}(E,R,N,\mu) &=& \int_0^1 dz \, z^{N-1} \frac{{\cal J}_{qg}(E,R,z,\mu)}{2(2\pi)^3} \nn \\
&=& \frac{\alpha_s C_F}{\pi} \left(   P_{gq}^N L + \frac{1}{2(N+1)}  + F_{N-2}-F_{N-1}+\frac{1}{2}F_{N} \right) \, ,
\eea
where $H_N$ is the harmonic number, $H_{N,2}$ is the generalized harmonic number of order 2, and 
$P_{ij}^N$, $F_N$, and $G_N$ are given by
\bea
F_N &=& \frac{2}{N+1}\left(-H_{N+1} + \sum_{j=1}^N \frac{1}{j \,2^j} -\log 2\right) , \\
G_N &=& \sum_{j=1}^{N} \frac{1}{j^2\, 2^j} - \sum_{k=1}^{N} \frac{1}{k}\left(  \sum_{j=1}^{k} \frac{1}{j \,2^j} - \log 2\right),
\\
P_{gg}^N &=& 2\left( -H_{N}+\frac{1}{N-1} -\frac{1}{N} +\frac{1}{N+1} -\frac{1}{N+2} \right) ,\\
P_{qq}^N &=&  -2 H_{N+1}+\frac{1}{N} + \frac{1}{N+1},\\
P_{gq}^N &=& \frac{N^2+N+2}{N(N^2-1)} ,\\
P_{qg}^N &=& \frac{N^2+N+2}{N(N+1)(N+2)} \,.
\eea
Note that 
\bea
F_N &=& {\cal O}\left( \frac{1}{N}\right), \nn \\ 
G_N &=& \frac{\pi^2}{12}  +  {\cal O}\left( \frac{1}{N}\right) \, , \eea
so in the large $N$ limit 
\bea
\tilde{{\mathcal J}}_{gg}(E,R,N,\mu) &=&1 + \frac{\alpha_s C_A}{\pi} \left( L_N^2 -\frac{\pi^2}{8}   +  {\cal O}\left( \frac{1}{N}\right) \right) \, , \\
\tilde{{\mathcal J}}_{qq}(E,R,N,\mu) &=&1 + \frac{\alpha_s C_F}{\pi} \left( L_N^2 -\frac{\pi^2}{8}  +  {\cal O}\left( \frac{1}{N}\right) \right) \, , 
\eea
where 
\bea
L_N = \ln \left(\frac{2 E \tan(R/2)}{N e^{\gamma_E}\mu}\right) \, . 
\eea
We see that the logarithms in $\tilde{\cal J}_{gg}(E,R,N,\mu)$ and  $\tilde{\cal J}_{qq}(E,R,N,\mu)$ are minimized at the scale
$2E\tan(R/2)/Ne^{\gamma_E}$. This is consistent with the expressions for ${\cal J}_{gg}(E,R,z,\mu)$ and ${\cal J}_{qq}(E,R,z,\mu)$
These logarithms are easily resummed using the jet anomalous dimension, however, we will not do this resummation 
in this paper as we compute moments with $N$ of order unity. Moments with $N=1$ are divergent because of the poles in $P_{gg}^N$, $P_{qg}^N$,
  $\tilde{\cal J}_{gg}$, and   $\tilde{\cal J}_{gq}$.


\end{document}